\documentclass[%
 reprint,
preprintnumbers,
nofootinbib,
 amsmath,amssymb,
 aps,
]{revtex4-2}

\usepackage[dvipdfmx]{graphicx}
\usepackage{amsmath}
\usepackage{amsfonts}
\usepackage{amssymb}
\usepackage{graphicx, rotating}
\usepackage{epsfig}
\usepackage{latexsym}
\usepackage{graphicx}
\usepackage{color}
\usepackage{amsmath,bm,amssymb}

\usepackage{color}
\usepackage[english]{babel}
\usepackage{hyperref}

\usepackage{textcomp}
\usepackage{ulem}

\newcommand{\Slash}[1]{{\ooalign{\hfil#1\hfil\crcr\raise.167ex\hbox{/}}}}

\newcommand{\beq}{\begin{equation}}  \newcommand{\eeq}{\end{equation}}
\newcommand{\bef}{\begin{figure}}  \newcommand{\eef}{\end{figure}}
\newcommand{\bec}{\begin{center}}  \newcommand{\eec}{\end{center}}

\newcommand{\vev}[1]{ \left\langle {#1} \right\rangle }

\newcommand{\SU}[1]{{\rm SU{#1} } }

\def\({\left(}
\def\){\right)}

\def\O{\mathcal{O}}
\def\U{\mathop{\rm U}}

\newcommand{\AND}{~{\rm and}~}
\newcommand{\EV}{ {\rm \, eV} }
\newcommand{\KEV}{ {\rm \, keV} }
\newcommand{\MEV}{ {\rm \, MeV} }
\newcommand{\GEV}{ {\rm \, GeV} }

\def\a{\alpha}

\def\f{\phi}

\def\l{\lambda}
\def\m{\mu}

\def\s{\sigma}

\def\D{\Delta}
\def\G{\Gamma}

\def\F{\Phi}

\def\*{\dagger}

\begin{document}

\preprint{TU-11??}
\title{
Singlet extensions and W boson mass in light of the CDF II result 
 }

\author{Kodai Sakurai}
 \email{kodai.sakurai.e3@tohoku.ac.jp}
\author{Fuminobu Takahashi}
 \email{fumi@tohoku.ac.jp}
\author{Wen Yin}%
 \email{yin.wen.b3@tohoku.ac.jp}
\affiliation{%
Department of Physics, Tohoku University, Sendai, Miyagi 980-8578, Japan
}%


\begin{abstract}
Recently, the CDF collaboration has reported 
the precise measurement of the W boson mass,
$M_W = 80433.5\pm 9.4 \MEV$, 
based on $8.8$ fb$^{-1}$ of $\sqrt{s}=1.96$ TeV $p\bar{p}$ collision data from the CDF II detector at the Fermilab Tevatron. 
This is about $7\sigma$ away from the 
Standard Model  prediction,
$M_{W}^{\rm SM}=80357 \pm  6 \MEV$.
Such a large discrepancy may be partially due to exotic particles  that radiatively alter the  relation between the W and Z boson masses. 
In this Letter, we study singlet extensions of the Standard Model focusing on the shift of the W boson mass. In the minimal extension with a real singlet field, using the bounds from the electroweak oblique parameters, B meson decays, LEP, and LHC, we find that the W boson mass shift is at most a few $\MEV$, and therefore it does not alleviate the tension between the CDF II result and the SM prediction. We then examine how much various bounds are relaxed when the singlet is allowed to decay invisibly, and find that the increase of the W boson mass does not exceed $5$\,MeV due to the bound from the Higgs signal strength. We also discuss phenomenological and cosmological implications of the singlet extensions such as the muon $g-2$ anomaly, axion/hidden photon dark matter, and self-interacting dark radiation as a possible alleviation of the Hubble tension. 
\end{abstract}

\maketitle

{\bf Introduction.--} Recently, the CDF collaboration has reported the precise measurement of the W boson mass~\cite{CDF:2022hxs} 
\beq M_W = 80433.5\pm 9.4 \MEV,\eeq based on  $8.8$ fb$^{-1}$ of $\sqrt{s}=1.96$ TeV $p\bar{p}$ collision data from the CDF II detector at the Fermilab Tevatron.
This is about  $80\MEV$ heavier than the 
Standard Model (SM) prediction~\cite{ParticleDataGroup:2020ssz}
\beq M_{W}^{\rm SM}=80357 \pm  6 \MEV, \eeq 
and the significance of the discrepancy is at $7\sigma$.
If true, it requires a light exotic particle beyond the SM, which radiatively alters the relation between the W and Z boson masses.
We note that the world average before the report was $80379\pm 12 \MEV $~\cite{ParticleDataGroup:2020ssz}, which is about $20$\,MeV heavier than the SM prediction, but is in stark tension with the CDF II result. 
Therefore, it is too early to conclude that the discrepancy between the CDF II result and the SM prediction is due to the beyond standard model (BSM) contribution. 
Still, it would be useful to see how difficult or easy it is  to have such a significant shift of the W boson mass by the BSM physics given the various experimental limits.

To deviate the W boson mass from the SM prediction, we need some non-decoupled BSM particles. Their masses should be around or lighter than the weak scale to change the relation between the W and Z boson masses or the electroweak precision parameter, $\D r$~\cite{Sirlin:1980nh}.
Since the exotic charged particles tend to be more tightly constrained\footnote{In the charged particle explanation, it was recently shown that when the wino mass is as small as $100\GEV$ the MSSM can give a significant contribution together with the explanation of the muon $g-2$~\cite{Bagnaschi:2022qhb}. 
Therefore it sheds light on the anomaly mediation scenario~\cite{Giudice:1998xp,Randall:1998uk} to explain the $g-2$~\cite{Yin:2021mls} (see also for the Higgs-anomaly mediaiton \cite{Yamaguchi:2016oqz, Yin:2016shg, Yanagida:2016kag,Yanagida:2018eho, Yanagida:2020jzy}, 
Higgs-gaugino mediation~\cite{Yamaguchi:2016oqz, 
Cox:2018vsv,
Agashe:2022uih} and some closely related topics~\cite{Chakraborti:2021dli,Endo:2021zal, Chigusa:2022xpq}) together with the correction of the W boson mass~\cite{Bagnaschi:2022qhb}.
That said, 
the pure wino enhancement of the W boson mass may be difficult due to the severe bound from the disappearing track, which requires the wino mass larger than $660\GEV$~\cite{ATLAS:2021ttq} (see also \cite{Sirunyan:2020pjd}). 
}, in this Letter, we study singlet extensions of the SM  focusing on the W boson mass shift. Such a gauge singlet could be a portal to the dark sector which involves the dark matter and/or dark radiation. Due to the gauge invariance, 
the dark sector could be coupled to the SM sector via the so-called Higgs portal interaction~\cite{Silveira:1985rk,Burgess:2000yq}. It was discussed that the loop contribution can induce a shift of the W boson mass via the mixing between the SM-like Higgs and the extra Higgs bosons~\cite{Lopez-Val:2014jva}. 

In this Letter, we first discuss the minimal singlet extension of the SM, and show that 
the LHCb bound for the B meson decay into K meson and a muon pair~\cite{LHCb:2012juf,LHCb:2015nkv,LHCb:2016awg}, the LEP bound~\cite{L3:1996ome} for the $e^- e^+\to s Z^{(\ast)}$, and the LHC bound on the Higgs signal strength require the lighter Higgs mass and the mixing with the SM-like Higgs boson to satisfy $m_s\gtrsim 5\GEV$ (for a moderately small mixing) and $\cos\alpha\lesssim 0.3$, respectively. 
We find that in this allowed region the singlet contribution to the W boson mass is at most a few $\MEV$. 
    We then consider the possibility that the singlet decays invisibly and find that the contribution to the W boson mass is slightly enhanced, but does not exceed $5$\,MeV. Interestingly, in this case, we derive a lower bound on the branching fraction of the invisible decay of the SM-like Higgs boson.  
    As a concrete model for such invisibly decaying singlets, we consider a complex scalar extension.  We will discuss the phenomenological and cosmological implications of the singlet extensions,
    such as the muon $g-2$ anomaly, axion dark matter, and self-interacting dark radiation.

{\bf The minimal singlet extension.--}
Let us consider the Higgs potential in the broken phase with the following form,
\beq 
V= m_{\rm mix}^2 \f_1 \f_2  + m_1^2 \frac{\f_1^2}{2}+  m_2^2 \frac{\f_2^2}{2}
\eeq 
with $\f_1 \AND \f_2$ being the neutral Higgs bosons in the SU(2) doublet and singlet, respectively. 
One can easily move to the mass eigenbasis $(s,h)$ as
\begin{align}\label{eq:basis}
\begin{pmatrix}
 s \\
 h \\
\end{pmatrix}=
R
\begin{pmatrix}
\f_1 \\
\f_2 
\end{pmatrix}
\end{align}
with a rotation matrix
 \begin{equation}
    \label{e:R}                                                                                           
          {R} = \begin{pmatrix}
                     \cos \alpha & \sin \alpha\\
            -\sin \alpha & \cos \alpha \end{pmatrix}.
 \end{equation}
Here $h$ is the SM-like Higgs boson with the mass $m_h=125.25\GEV$, and
we assume that $s$ is lighter than $h$ and has the mass $m_s$. 
 Thus, we have two free parameters, $m_s$ and $\cos\a$.  When  $\cos\a$ is small, it is approximately given by $\cos\alpha \simeq m_{\rm mix}^2/|m_1^2- m_2^2|.$

\begin{figure}[!t]
\begin{center}  
  \includegraphics[width=.43\textwidth]{ 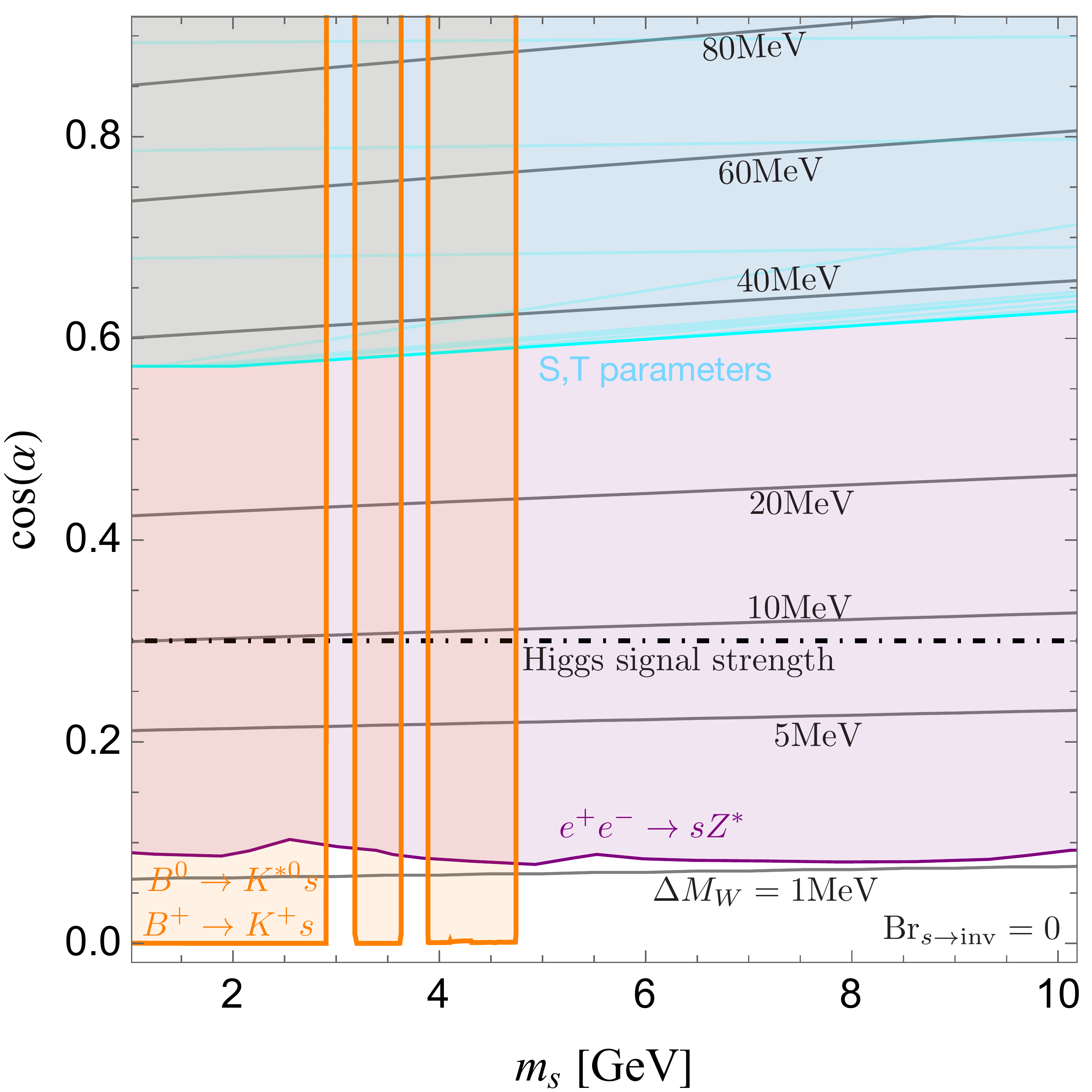}
    \includegraphics[width=.43\textwidth]{ 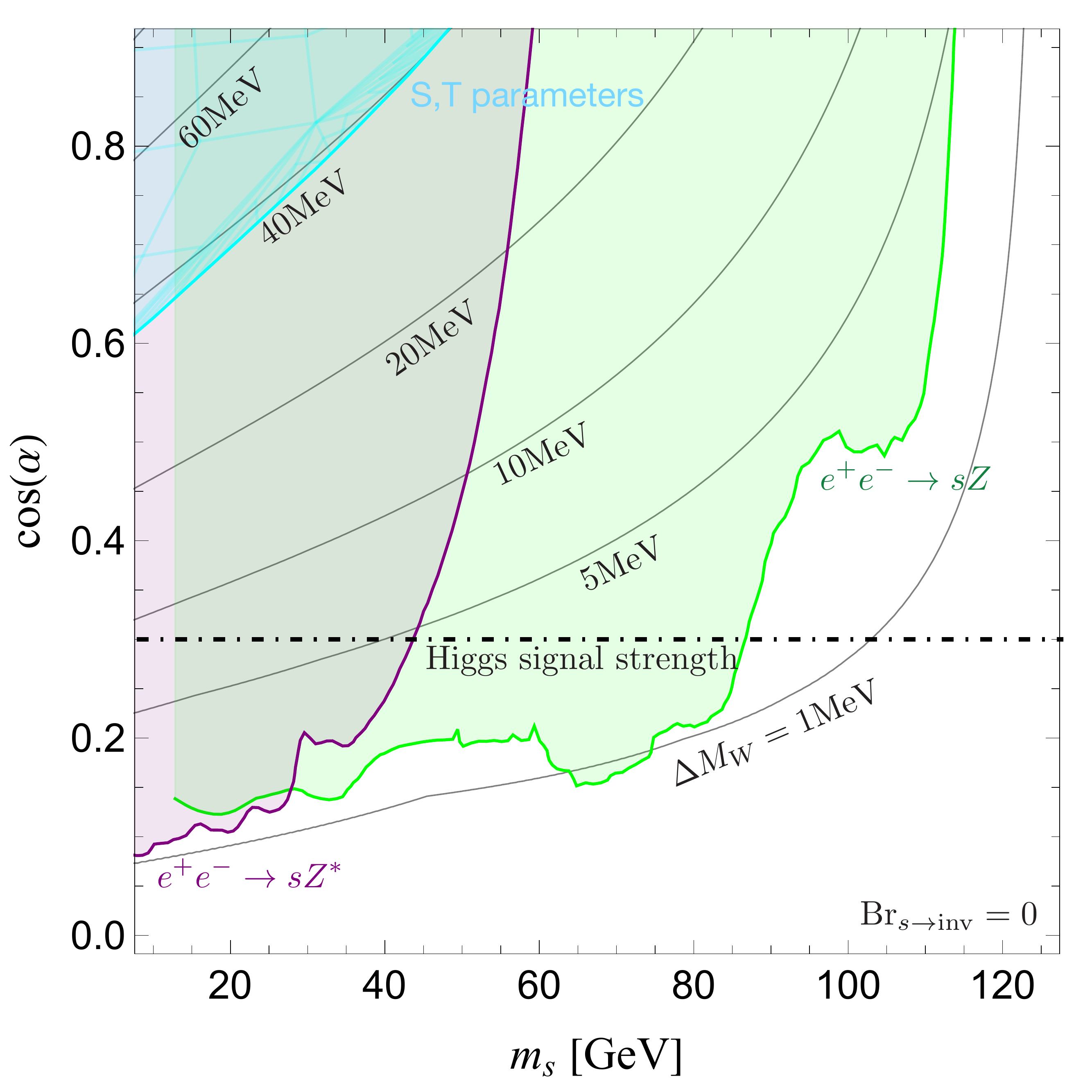}
      \end{center}
\caption{
The contour of $\D M_W \equiv  M_W^{\rm SM+singlet}-M_{W}^{\rm SM}$ on the $(m_s,\cos\a)$ plane in the single singlet extension of the SM for $m_s < 10$\,GeV (top) and $m_s > 10$\,GeV (bottom). Also shown are the bounds of the oblique parameters $S$ and $T$ (light blue),
LHCb~\cite{LHCb:2012juf,LHCb:2015nkv,LHCb:2016awg} (orange) and LEP~\cite{L3:1996ome,LEPWorkingGroupforHiggsbosonsearches:2003ing} (purple and green). The horizontal dot dashed line represents the upper bound on $\cos \alpha$ coming from Higgs signal strength~\cite{Aad:2019mbh, CMS:2020gsy}.
} \label{fig:1} 
\end{figure}

When $s$ is lighter than $h$, its loop contribution enhances the 
mass of the W boson~\cite{Lopez-Val:2014jva}. 
We show in Fig.\ref{fig:1}
the contours of $\D M_W \equiv  M_W^{\rm SM+singlet}-M_{W}^{\rm SM}$ in the
$(m_s,\cos\a)$ plane, where $M_W^{\rm SM+singlet}$ ($M_{W}^{\rm SM}$) denotes the W boson mass in the singlet extension of the SM (the SM). 
This is calculated by utilizing {\tt H-COUP}~\cite{Kanemura:2017gbi,Kanemura:2019slf} with modification.
We evaluate the mass of W boson by
\begin{align}
M_W=\frac{M_Z}{2}\left(1+\sqrt{1-\frac{4\pi\alpha_{\rm em}}{\sqrt{2}G_F M_Z^2(1-\Delta r)}}\right),
\end{align}
choosing $M_Z$, $G_F$, and $\alpha_{\rm em}$ as inputs for the electroweak (EW) parameters.  
In Fig.\ref{fig:1}, the bound from the oblique parameters $S$ and $T$~\cite{Peskin:1991sw} (light blue) {with the fitting values in Ref.~\cite{Haller:2018nnx}}, the LHCb bound (orange)~\cite{LHCb:2012juf,LHCb:2015nkv,LHCb:2016awg}, and the LEP bounds (purple and green)~\cite{L3:1996ome,Winkler:2018qyg,LEPWorkingGroupforHiggsbosonsearches:2003ing} are also shown. One can see that the bound from the $S$ and $T$ parameters exclude the region with $\Delta M_W \gtrsim 40$\, MeV, which makes it difficult for this model to explain the CDF II result. 
 The LEP bounds consist of the two analyses (purple and green) where the singlet is assumed to decay into the SM particles via the mixing. The bound reads $\cos \alpha \lesssim 0.1$ for  $m_s \lesssim 10\GEV$. This  constrains $\Delta M_W$ to be about $1$\,MeV or less.
On the other hand, when the mass is larger than $10\GEV$, the bound on the mixing gradually becomes weaker with $m_s$, and the $\D M_W$ can be slightly larger. However, there is another bound on the mixing from the measurement of the Higgs signal strength at the LHC~\cite{Aad:2019mbh, CMS:2020gsy}, which constrains $\cos \alpha \lesssim 0.3$,
{where we use the lower limit for the scaling factor for Z boson $ k_Z=\sin \alpha \gtrsim 0.95.$ We note that this upper bound for $\cos\alpha$ changes if $m_s$ is close to $m_h$~\cite{Robens:2015gla}.}
As a result, $\D M_W$ is at most a few MeV in the region satisfying all these experimental bounds.
For instance, for {$m_s=90\GEV$ and $\cos\a=0.3$, we get $\D M_W=1.6~{\rm MeV}$}.
Note that, if one considers the case in which the two neutral Higgs bosons are almost degenerate~\cite{Abe:2021nih, Cho:2021itv, Sakurai:2022cki},
this bound of $\cos \a$ can be relaxed. However, the contribution to $\D M_{\rm W}$ is suppressed by the mass difference.\footnote{Many almost degenerate singlets mixed with Higgs boson also do not make things better since only one combination of the scalars contributes to the W/Z boson masses.}

{\bf Invisible decays of the singlet and SM-like Higgs.--} 
The previously shown LEP and LHCb bounds assume that the $s$ decays visibly via the mixing with $\f_1$.
One possible way to relax the experimental bounds is to assume that $\f_2$ is coupled to a dark sector and it decays into the invisible particles. 
In such models the singlet is a portal to the dark sector which may contain dark matter particles or other particles that are relevant to the muon $g-2$ anomaly, the Hubble tension, etc.

In fact it is non-trivial if the bounds can be relaxed by including the invisible decay of the singlet; for 
$m_s\lesssim 30\GEV$, the LEP bound on the mixing for the invisible decay is more stringent than that for the visible decay~\cite{L3:1996ome}.
On the other hand, when $m_s\gtrsim \O(10)\GEV$,  the constraint from $e^+ e^-\to Z s $ followed by  
the $s$ invisible decay is weaker than that for the visible decay~\cite{LEPWorkingGroupforHiggsbosonsearches:2003ing,DELPHI:2003azm}. 
Thus one can relax the bound to expand the allowed parameter region
for $m_s\gtrsim \O(10)\GEV$ by introducing the invislbe decays.

\begin{figure}[!t]
\begin{center}  
  \includegraphics[width=.35\textwidth]{ 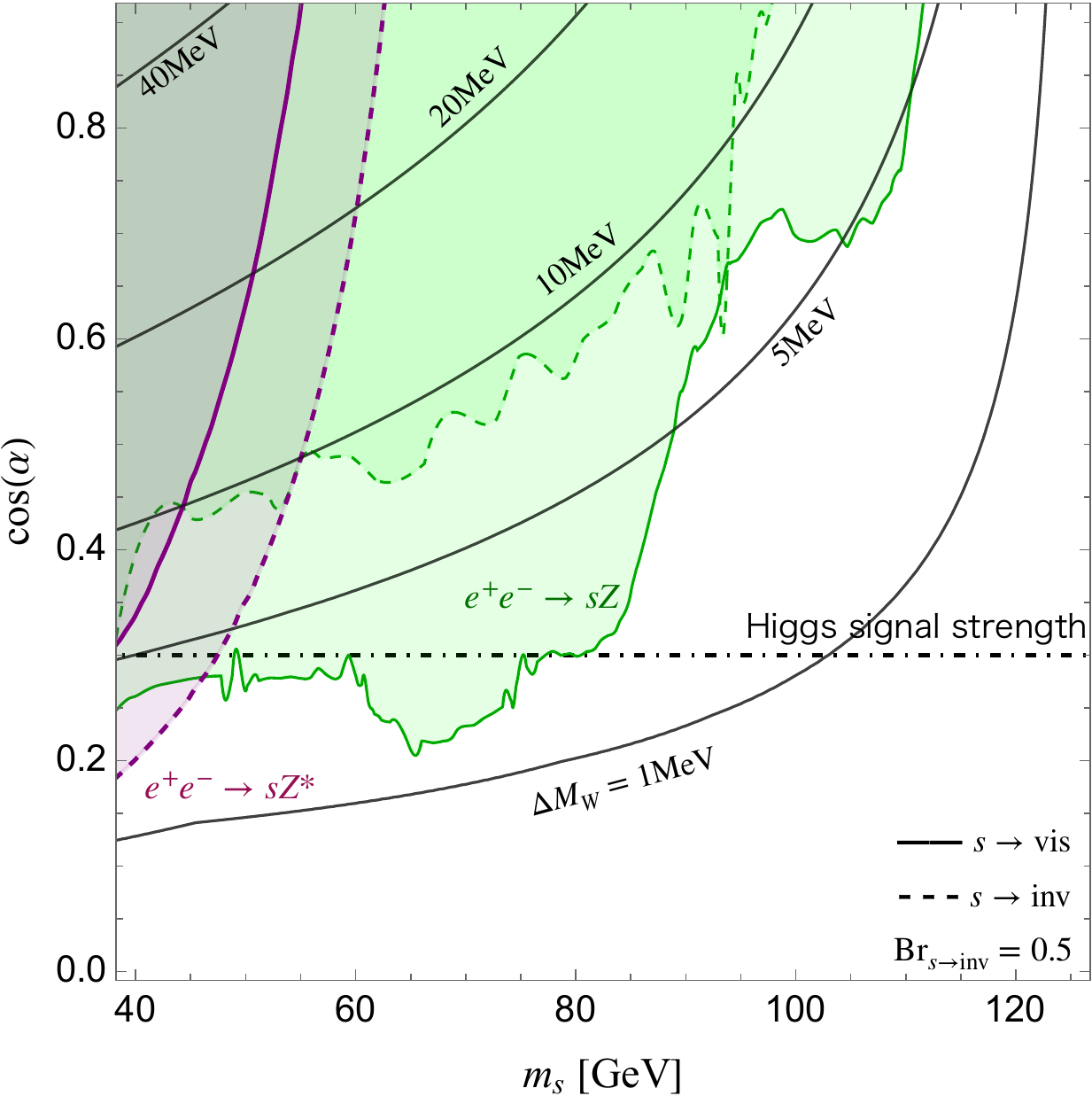}
  \includegraphics[width=.35\textwidth]{ 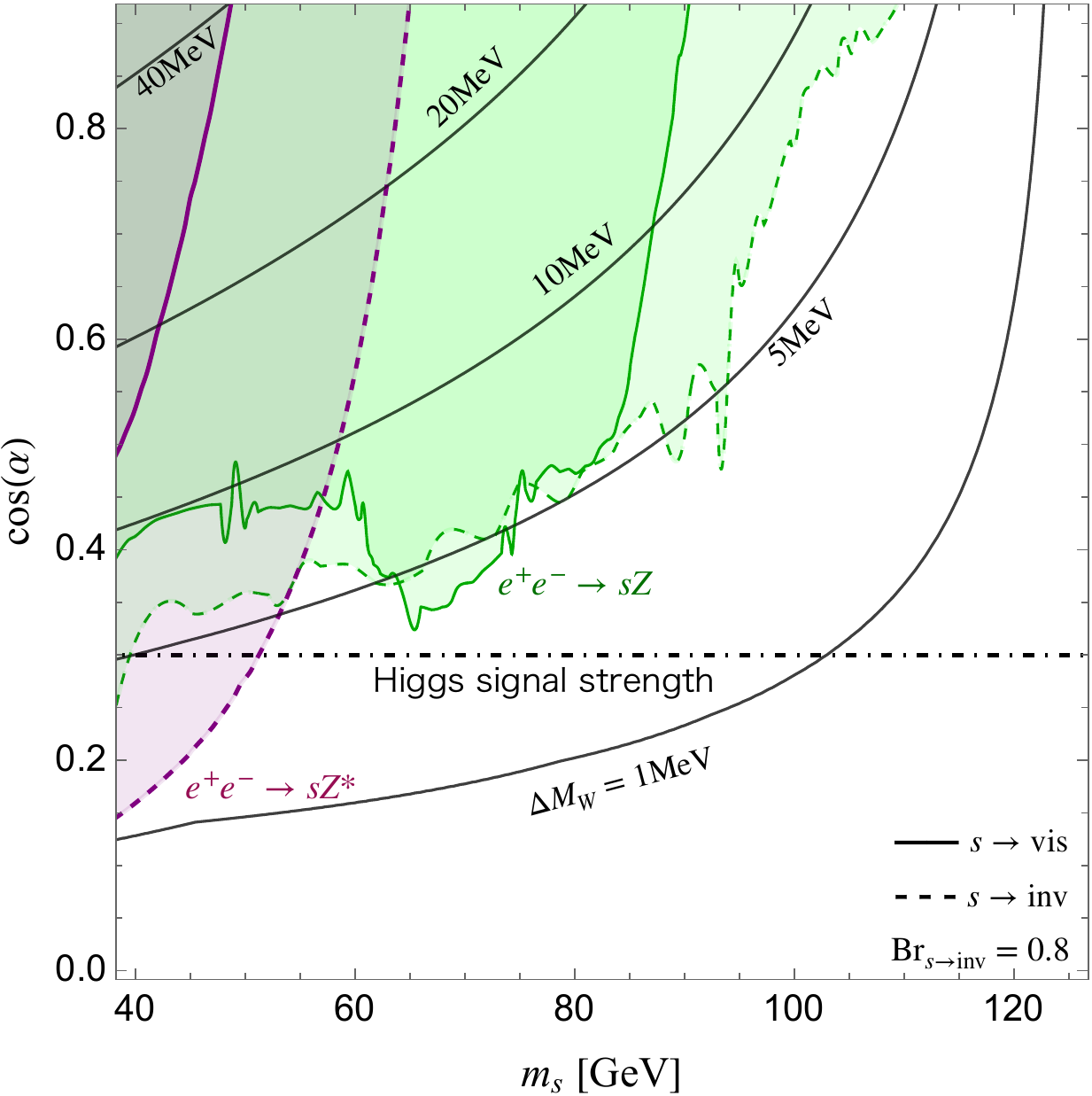}
  \includegraphics[width=.35\textwidth]{ 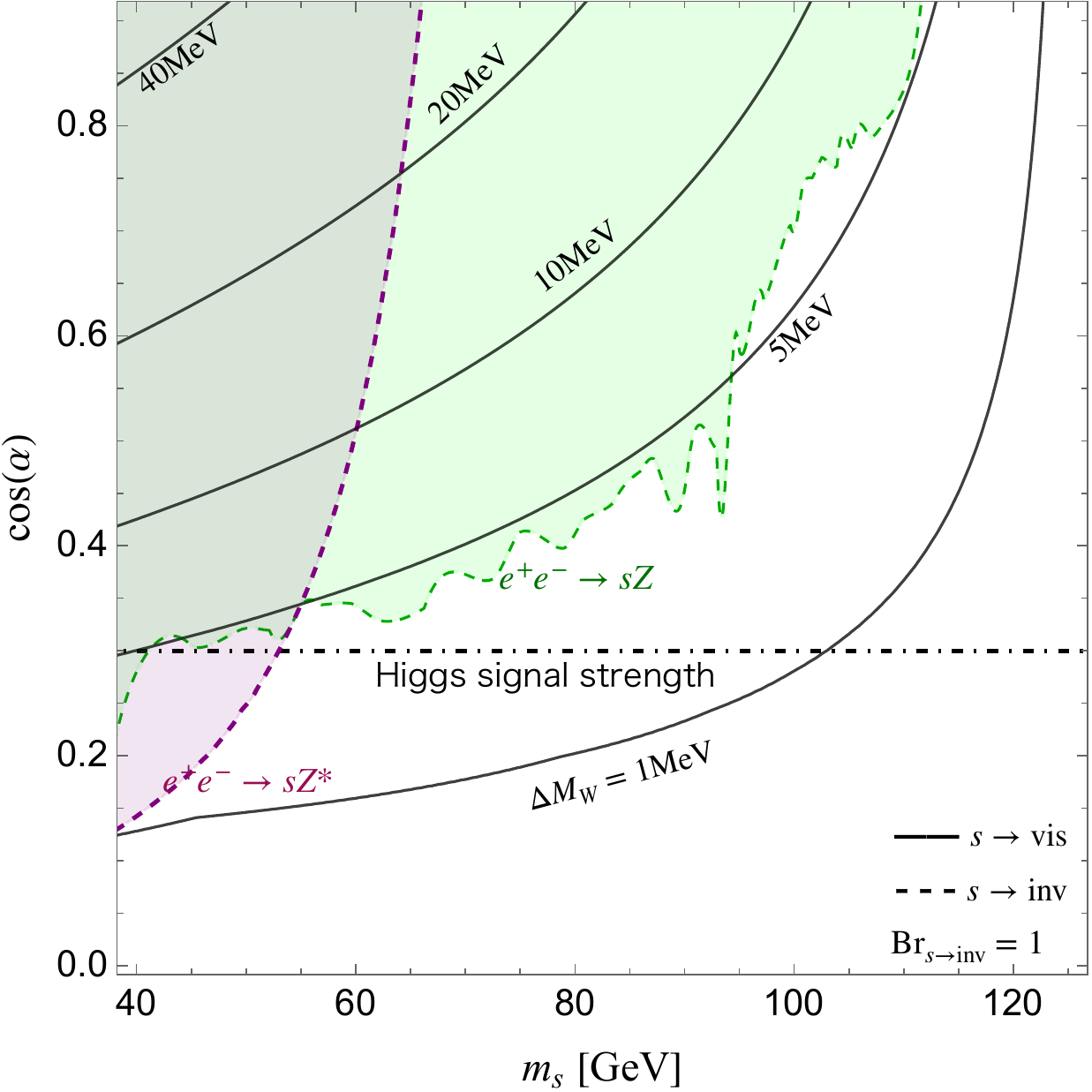}
      \end{center}
\caption{
The contours of $\D M_W \equiv  M_W^{\rm SM+singlet}-M_{W}^{\rm SM}$ in the $(m_s,\cos\a) $ plane in the dark sector extension of the SM.
The singlet decays into dark particles with the branching fraction ${\rm Br}_{s \to {\rm inv}}=0.5,0.8$ and $1$ from top to bottom. 
The green shaded region bounded by solid (dashed) line is from $e^+ e^-\to Z s $ with $s $ to visible (invisible) particles from the LEP data translated from Ref.\,\cite{LEPWorkingGroupforHiggsbosonsearches:2003ing} (\cite{DELPHI:2003azm}). 
The purple bounds correspond to $e^+ e^-\to Z^* s $~\cite{L3:1996ome}
} \label{fig:2} 
\end{figure}

We show  the viable parameter region in Fig.\ref{fig:2} 
by assuming a non-vanishing decay rate of $s$ to dark particles, $\G_{s\to {\rm inv}}\neq 0$. The branching fraction of the invisible decay rate of $s$ is assumed to be 
${\rm Br}_{s\to \rm inv}=0.5,0.8,1$ from top to bottom panels. 
Here, we can approximately estimate the branching fraction as
\beq {\rm Br}_{s\to \rm inv}\simeq \frac{\G_{s\to \rm inv}}{\G_{s\to \rm inv}+\cos^2\a \frac{m_s}{m_h} \G_{h}^{\rm SM}}\eeq 
where {$\G_h^{\rm SM}$ is the  Higgs decay rate  in the SM,
and it can be approximated by the partial decay rate to $b \bar{b}$ 
as long as $m_s$ is lighter than $m_h$.}
As one can see from Fig.\ref{fig:2}, if ${\rm Br}_{s\to \rm  inv}\gtrsim 0.8$, the bound on $\cos\a$ is relaxed, and $\D M_{W}$ can be slightly  larger. Note however that the Higgs signal strength bound is still stringent and the effect is not drastic; for instance, we obtain $\D M_W =3.8\MEV$ at $m_s=55\GEV$ and $\cos\alpha=0.3$.

Since there is a mixing, the presence of the $s$ invisible decay implies that $h$ also decays invisibly via the mixing. (There could also be other model-dependent invisible decay channels as we will discuss.) 
This process can be treated in a more-or-less model independent way if
we express the SM-like Higgs invisible decay rate as
\beq 
\G_{h\to \rm inv}= \cos^2\a \(\frac{m_h}{m_s}\)^n \G_{s\to \rm inv},
\eeq 
where $n$ depends on the decay process. 
For instance, if the $s$ decays into a pair of light scalars, $a$, via the cubic term $\f_2 aa,$
we have $n= -1$.
When $s$ decays into a pair of light fermions via a Yukawa interaction, we have $n=1$. 
Interestingly, one finds that the invisible decay fraction of the SM-like Higgs satisfies\footnote{ In the numerical estimation, we take account of the branching fraction of the Higgs boson to $b\bar{b}$. }
\begin{align}
{\rm Br}_{h\to \rm inv}&\simeq \frac{\G_{h\to \rm inv}}{\G_{h\to \rm inv}+\G_{h}^{\rm SM}} \\ 
&=\frac{1}{1 + \frac{(1-{\rm Br}_{s\to \rm inv})\left(\frac{m_h}{m_s}\right)^{1-n}}{  \cos\a^4 {\rm Br}_{s\to\rm inv}}}.
\end{align} 
Thus, when we consider a relatively large mixing, e.g. $\cos \a\sim 0.3$, to enhance the $\D M_W$,  the  invisible decay of the SM-like Higgs is predicted
to be 
${\rm Br}_{h\to \rm inv}>0.1\% (1\%)$ 
for ${\rm Br}_{s \to {\rm inv}}\gtrsim 0.8$ with $n=-1$ (1) for $120\GEV>m_s>50\GEV$. I we increase $n$, the lower bound of ${\rm Br}_{h\to \rm inv}$ mildly increases.
Therefore this scenario can be tested not only by further measuring the Higgs boson coupling, but also by the invisible decay in the future  
 lepton colliders such as FCC-ee, CEPC, ILC, CLIC, and muon colliders~\cite{Asner:2013psa,dEnterria:2016sca,Abramowicz:2016zbo,CEPC-SPPCStudyGroup:2015csa, Ankenbrandt:1999cta, Delahaye:2019omf}, which will be able to reach ${\rm Br}_{h \to \rm inv}>0.1\%$ level.

 {\bf A concrete dark sector model.--}
 In the following we study the case of $n=-1$, 
a concrete example, a complex scalar extension of the SM, which incorporates the invisible decay. 

The potential in the symmetric phase is given as\footnote{A slight modification of the model has been studied in the context of axion/dark photon dark matter production via the phase transition~\cite{Nakayama:2021avl},
 the UV model of a CP-even ALP~\cite{Sakurai:2021ipp}, 
WIMP~\cite{Barger:2008jx, Barger:2010yn, Gonderinger:2012rd}\cite{Ishiwata:2018sdi, Cline:2019okt, Grzadkowski:2020frj, Abe:2020iph, Abe:2021nih}, electroweak baryogenesis~\cite{Barger:2008jx,Cho:2021itv} and collider physics~\cite{Chen:2019ebq, Grzadkowski:2020frj, Abe:2021nih, Bhattacherjee:2021rml, Sakurai:2021ipp,Sakurai:2022cki}.}
\beq
V=-m^2_\F|\F|^2+{\lambda} |\F|^4 +\l_P |h|^2 |\F|^2 + \lambda_H |H|^4-\mu_H^2 |H|^2.  \label{V}
\eeq
Here $\F$ and $H$ are  the dark and SM Higgs field, respectively. $\F (H)$ is  gauge singlet (doublet), whose vacuum expectation value (VEV) will spontaneously break the $\U(1)$ global (the $\SU(2)_L\times \U(1)_Y$ gauge) symmetry.
$\lambda_P, \l (>0) \AND \l_H (>0)$, $\m_H^2 (>0) \AND m_\F^2(>0)$ 
are the model parameters.

The $\Phi$ and $H$ develop nonzero VEVs, $\vev{\F}=v_s$ and $\vev{H_0}=v$,
and the corresponding symmetries get spontaneously broken. Since $\U(1)$ is a global symmetry, 
there is a massless Nambu-Goldstone boson (NGB), $a$.
The NGBs of $\SU(2)_L\times \U(1)_Y$ are eaten by the weak gauge bosons. 
Namely, we have
\begin{align}
H=
\begin{pmatrix}
G^{+} \\
\frac{1}{\sqrt{2}}(v+\f_1+iG^0)
\end{pmatrix},\quad
\Phi=\frac{1}{\sqrt{2}}(v_{s}+\f_2+ia),
\end{align}
where $G^{+}$ and $G^{0}$ are the NGBs that are eaten by the longitudinal mode of the weak gauge bosons. 
Here $a$ is the NG boson of $\U(1)$. 
We can give mass to $a$ via an explicit breaking of $\U(1)$\footnote{
If the axion potential has a non-trivial domain wall number, 
it may cause a serious domain wall problem.  
If we want the axion to be dark matter,
 we need to suppress the CP violation of the explicit breaking of $\U(1)$ or the mixing between the dark gauge boson with the photon~\cite{Sakurai:2021ipp}, since otherwise $a$ is not long-lived enough to be dark matter. }, or gauging $\U(1)$. Then we get the massive axion or dark photon, which can have interesting implications such as the Hubble tension, dark matter, muon $g-2$, as we will mention shortly.

 In this model $s$ decays into a pair of $a$,
 \beq s\to aa\eeq 
  via a rate of 
 \beq
  \Gamma_{s\to \rm inv} \sim  \frac{m^3_s}{32\pi v^2_s} ,
 \eeq
 where we neglect the mass of the NGB $a$ and assumed $\cos \a \sim \lambda_P v_s v/m_h^2 \lesssim 0.3$.

$h$ can also decay into a pair of $aa$ via the coupling of $\lambda_P v \sin \a h a^2/2$, which is not included in the previous discussion. The decay rate to $aa$ can be well approximated by\footnote{Due to the additional contribution, it corresponds to $n=3$ but not $-1$.}
\beq 
 \Gamma_{h\to aa} \simeq   \frac{m_h^3}{32\pi v_s^2}\cos^2\a.
\eeq 
We will not consider $h \to ss$ since in the region of interest it is kinematically suppressed. 

The invisible decay rate of the SM-like Higgs boson needs to satisfy ${\rm Br}_{h\to \rm inv}\lesssim 0.1$~\cite{ATLAS-CONF-2020-052} (see also Refs.~\cite{ATLAS:2019cid, CMS:2018yfx}). 
For $\cos\alpha=0.3, {\rm Br}_{s\to \rm \rm inv}\sim 0.8$ we obtain $ Br_{h\to \rm inv}= 0.01-0.1$ if $50\GEV < m_s < 120\GEV$, which can satisfy the bound.
In this case, by taking ${\rm Br}_{s\to \rm inv}>0.8$, we derive 
\beq 
v_s \lesssim (100-1000)/\cos\a,~~ \lambda_P \gtrsim (0.001-0.1) \cos^2\a.
\eeq 
In the following, by taking this condition with $\cos\a\sim \O(0.1)$, we study some phenomenological and cosmological implications.

{\bf Muon $g-2$ from {$L_\mu-L_\tau$} gauge symmetry.--}
One can gauge the $\U(1)$ as the $L_\mu- L_\tau$ symmetry.
The interesting point is that the gauge boson mass is 
\beq 
m_{Z'}\sim 100\MEV \(\frac{g'}{10^{-4}}\) \(\frac{v_s}{1000\GEV}\)
\eeq 
which $g'$ is the gauge coupling. 
This is in the favored range to explain the muon $g-2$ anomaly.
The muon $g-2$ anomaly confirmed in the Fermilab experiment\,~\cite{Muong-2:2021ojo}, shows a 4.2 $\s$ deviation from the SM prediction~\cite{Roberts:2010cj, Davier:2017zfy,Keshavarzi:2018mgv, Keshavarzi:2019abf,  Aoyama:2020ynm, Chao:2021tvp}\footnote{See also Ref.\,\cite{Borsanyi:2020mff} and \cite{Crivellin:2020zul, Keshavarzi:2020bfy}.}
 \beq
 \D a_\mu= (25.1 \pm 5.9)\times 10^{-10}. 
 \eeq
  The contribution from the $L_\mu-L_\tau$ can be estimated as ~\cite{Fayet:2007ua, Pospelov:2008zw,Ilten:2018crw, Bauer:2018onh} (see also the bounds in the references)
 \beq
\D a^{Z'}_\mu\approx \frac{q^2 g'^2}{8\pi^2} \int_{0}^1 dz \frac{2 m_\mu^2 z (1-z)^2}{m_\mu^2 (1-z)^2+m_{Z'}^2 z}
 \eeq
 at the 1loop level where $q$ is the charge of the leptons in the unit of the $\f_2$ charge. We may need $m_{Z'}<2m_\mu$ to evade the bound from the displaced vertex of muons from the SM Higgs decay. With $q=\O(1)$ we can explain the muon $g-2$ anomaly.

{\bf Cosmological implications.--} Let us discuss the physics of $a$ (which may also denote the dark photon that eats it). 
If the reheating temperature of the Universe is too high, $\gtrsim 100\GEV$, $a$ is thermalized due to the large $\l_P$. Thus it either overcloses the Universe with $m_a\gg \EV$ or it becomes the hot dark matter. 
Therefore we need to consider the low-reheating temperature for it being the cold dark matter with $m_a\gtrsim 20\KEV$ e.g.~\cite{Sakurai:2021ipp}.

We may take the $m_a\to 0$ to have the dark radiation~\cite{Weinberg:2013kea}. The decoupling temperature of $a$ is then around $m_s/20\gtrsim \O(1\GEV)$ for $\O(10)$ GeV $s$. This leads to a deviation in the effective neutrino number
$\D N_{\rm eff}\sim 0.07-0.08 (0.03-0.04)$ for  a gauge boson (NGB) e.g.~\cite{Baumann:2017gkg}. In particular, when the dark radiation with $\D N_{\rm eff}= \O(0.1)$ is self-interacting~\cite{Jeong:2013eza} the Hubble tension can be alleviated~\cite{Planck:2018vyg,Riess:2011yx, Bonvin:2016crt,Riess:2016jrr, Riess:2018byc,Birrer:2018vtm,Riess:2021jrx}. (See, also, Refs.~\cite{Freedman:2020dne,Khetan:2020hmh}, and reviews ~\cite{Freedman:2017yms, Verde:2019ivm,DiValentino:2020zio,Schoneberg:2021qvd} for the Hubble tension.) 
Indeed, if we extend $\U(1)$ to $\SU(N)$ gauge group with $\f_2$ being a fundamental representation Higgs~\cite{Jeong:2013eza}, we get a spontaneous breaking of $\SU(N)\to \SU(N-1).$ Thus there are $((N-1)^2-1)$ massless and $2N-1$ massive gauge bosons. The massive gauge boson is charged and induces the higher dimensional coupling between $s$ and the massless ones, with a higher dimensional coupling $g'^2/(16\pi^2v_s)$. This generates the self-interacting dark radiation with $\D N_{\rm eff}\sim \O(0.1)$ for $N=3-4$~\cite{Jeong:2013eza}.\footnote{The charged heavy gauge boson is stable, although the abundance can be small. The cosmology of the system should be interesting.}

{\bf Discussions.--} So far we have considered various extensions to the SM. In particular, a simple dark sector model can induce $\D M_W$ close to 5~MeV. 

Let us comment on the possibility of more {complex} dark sector to further enhance the W boson mass shift. The model is ad hoc, but it will provide an example to enhance $\Delta M_W$. We will introduce many particles to radiatively vary the couplings of $s$ and $h$ to the SM particles.
We would like to have such a scale dependence of the couplings that the $h$ coupling is enhanced to evade the signal strength bound but $s$ coupling is suppressed to evade the LEP bounds. 
To this end, we may consider that $\f_2$ couples to gauge singlet fermions and bosons. 
This gives wave function corrections to the kinetic term of $\f_2$ and thus to the $h,s$ couplings.
The radiative correction from 
a fermion (not too light boson) loop induces a positive (negative) correction to the coupling between $h$ and the SM particles. 
In addition, the particles much lighter than 125GeV, may induce a negative correction solely to the coupling between $s$ and SM particles. 
However, from the perturbative unitarity, we expect that $\D M_W$ cannot be too larger than $10\MEV$.\footnote{Another possibility may be that a scalar field couples to the Higgs potential in such a way that its expectation value changes the Higgs mass with VEV kept intact. 
When a non-trivial configuration of the scalar field, e.g. domain wall, goes through Earth, the Higgs mass may be decreased. 
The radiative correction enhances the W boson mass shift. A certain domain wall width may explain the discrepancy between different experimental results. 
}

In summary, we have examined the loop contribution to the W boson mass in the singlet extensions of the SM, taking account of various experimental bounds. We have found that $\Delta M_W$ is at most a few MeV in the minimal singlet extension, and it can be slightly increased by allowing the invisible decays of the singlet, but it does not exceed $5$\,MeV especially due to the tight bound on the mixing angle coming from the Higgs signal strength.
Therefore, we conclude that the contribution to the W boson mass is limited in the singlet extensions and cannot explain the large discrepancy between the CDF II result and the SM prediction, while it can partially relax the mild tension between the other experiments and the SM prediction.

{\bf Acknowledgement.--}
This work was supported by JSPS Core-to-Core Program (grant number: JPJSCCA20200002) (F.T.), JSPS KAKENHI Grant Nos.  20H01894 (K.S. and F.T.) 20H05851 (F.T. and W.Y.), 21K20363 (K.S.), 21K20364 (W.Y.), 22K14029 (W.Y.), and 22H01215 (W.Y.).

\clearpage
\appendix

\bibliography{Wboson}
\end{document}